\documentclass[epj]{svjour}
\usepackage{graphicx}
\usepackage{dcolumn}

\usepackage{amsmath}
\usepackage{amssymb}
\usepackage{bm}
\usepackage{epsfig, graphicx, euscript}

\newcommand{\beq}{\begin{equation}}
\newcommand{\eeq}{\end{equation}}
\newcommand{\bea}{\begin{eqnarray}}
\newcommand{\eea}{\end{eqnarray}}
\newcommand{\eps}{\varepsilon}
\newcommand{\bfg}{\boldsymbol}

\begin{document}

\title{Magnetic moments of odd-odd spherical nuclei}

\author{0.\,I. Achakovskiy\inst{1}  \inst{2} \and S.\,P. Kamerdzhiev \inst{2}
\fnmsep\thanks{Corresponding author; \email{kamerdzhiev@ippe.ru}}
\and E.\,E. Saperstein\inst{3} \and S.\,V. Tolokonnikov
\inst{3},\inst{4}
}
\institute{ Institute for Nuclear Power Engineering, 249030 Obninsk,
Russia \and Institute for Physics and Power Engineering, 249033
Obninsk, Russia\and Kurchatov Institute, 123182, Moscow, Russia \and
Moscow Institute of Physics and Technology, 141700 Dolgoprudny,
Russia}

\abstract{ Magnetic moments of more than one hundred  odd-odd
spherical nuclei in ground and excited states are calculated within
the self-consistent TFFS based on the EDF method by Fayans {\it et
al}. We limit ourselves to  nuclei with a neutron and a proton
particle (hole) added to the magic or semimagic core.   A simple
model  of no  interaction between the odd
nucleons is used.  In most the cases we analyzed, a good agreement
 with the experimental data is obtained. Several cases are considered
 where this simple model does not work and it is necessary to go beyond.
The unknown values of magnetic moments of many unstable odd and
odd-odd nuclei are predicted including sixty values  for excited
odd-odd nuclei.}


\authorrunning{0.I. Achakovskiy et al.}
\titlerunning{Magnetic moments of odd-odd spherical nuclei}

\maketitle

\section{Introduction}
Description of nuclear magnetic moments  was always some sort of
polygon for nuclear theory. In addition, a systematic analysis of
magnetic moments gives a unique information about spin components of
the effective nuclear forces. Modern Radioactive Ion Beam facilities
provide access to long chains of isotopes and isotones including the
radioactive ones. Spectroscopy techniques using high-intensity
lasers allow for precision measurements of  spins and magnetic
moments of odd and odd-odd nuclei in their ground and isomeric
states. As a result, there is now a huge number of experimental data
for magnetic moments \cite{stone,BNL} of stable and unstable nuclei.
A consistent theoretical description of these data is necessary for
safe interpretation of the structure of the ground and excited
states of odd and odd-odd nuclei. As an example we mention an
observation of Refs.  \cite{mu1,mu2} that a sharp change of the
magnetic moment value between neighboring odd isotopes may give a
signal of the deformation phase transition.

Last decade, magnetic moments of nuclei of $1f-2p$ shell, $40<A<90$,
were successfully described within the Many-Particle Shell Model
(MPSM) \cite{BABr,Q_Cu}. This approach is very comprehensive as it
takes into account different many-particle correlations. The
necessity to introduce many parameters  of the effective
interaction, the single-particle mean field and the effective
particle charges is some deficiency of the MPSM. These parameters
are valid for the shell under consideration only and should be
changed provided one goes to any other shell. In addition, the
domain of the MPSM applications is limited up to now to nuclei with
$A<90\div100$.

To our knowlege,  the Theory of Finite Fermi Systems (TFFS)
\cite{AB1}  was the only consistent approach used for systematic
analysis of magnetic moments in heavy nuclei. It operates with
rather small number of the Landau--Migdal (LM) parameters and those
for the quasiparticle local charges $e_q$ universal for all nuclei,
except for the lightest ones.

In Refs. \cite{mu1,mu2} cited above, a wide array of magnetic
moments of spherical odd magic and odd semimagic nuclei
\footnote{For brevity, we name odd nuclei with one particle or hole
added to the even magic core as
odd nearmagic nuclei.
 Analogously, we say ``odd semimagic'' nuclei provided the even core is semimagic.}
was successfully described within the self-consistent TFFS based on
the Generalized Energy Density Functional(EDF) by Fayans {\it et
al.} \cite{Fay1,Fay5,Fay} with  the DF3 set of parameters
\cite{Fay5,Fay} to calculate the self-consistent mean field. The
spin-dependent LM parameters were used found many years ago
\cite{mu0} from the analysis of various magnetic characteristics of
$^{208}$Pb and its odd neighbors. As to the local charge $e_q$, a
tensor term with the parameter $\zeta_t$ was added to the usual TFFS
local charge with the spin and orbital parameters $\zeta_s$ and
$\zeta_{\,l}$. Correspondingly, the $\zeta_s$ value was a little
change in comparison with \cite{mu0} but the $\zeta_{\,l}$ parameter
remained the same. The so-called ``single-quasiparticle
approximation'' was used based on the consideration of an odd
nucleus in terms of  one quasi-particle added to the even-even magic
or semimagic core in the fixed state $\lambda=(n,l,j,m)$ with the
energy $\varepsilon_{\lambda}$.
 According to TFFS \cite{AB1},  a quasi-particle  differs  from
 a particle of the single-particle model in two respects.
First, it possesses the local charge $e_q$  and, second, the core is
polarized
 due to the interaction between the particle and the core nucleons via the
 LM effective interaction.
 In other words, the quasiparticle possesses the
 effective charge $e_{\rm eff}$ caused by the polarizability  of the core which is found
 by solving the TFFS equations. In the MPSM, it is introduced as a fitted parameter for
 the shell under consideration.
Experimental data on magnetic moments of more than one hundred odd
nuclei were reproduced, with several exceptions, with an accuracy of
$(0.1\div 0.2) \mu_N$, $\mu_N$ being the nuclear
magneton.
 However, there are several  odd-proton nuclei
where the accuracy was essentially worse \cite{mu1}. E. g., in the
magic $^{209}$Bi nucleus the deviation from experiment reaches
$\simeq 0.4 \mu_N$ and in the semimagic $^{125,127}$In, even $\simeq
0.5 \mu_N$.

Recently, a similar approach has been successfully applied  for the
quadrupole moments $Q$ of magic and semimagic odd nuclei
\cite{Q} with an improved version\\ DF3-a \cite{Tol-Sap} of the
original DF3 functional \cite{Fay5,Fay} which  differs from DF3 only
a little, with the spin-orbit and effective tensor force parameters
of the EDF. For this problem, the spin-independent LM amplitude
enters the TFFS equations for $Q$ which can be found directly as the
second variational derivative over density of the EDF used. Again
rather good accuracy in description of the experimental quadrupole
moments, $\simeq (0.1\div 0.2)$ b, was obtained, and again in
several cases the agreement was essentially worse.

These cases of disagreement for both the magnetic  moments and
quadrupole ones  indicated that sometimes the simplest
one-quasiparticle model works not sufficiently well and some
corrections should be taken into account. The phonon coupling (PC)
effects are the most evident candidate for such corrections. Very
recently, they were analyzed for magnetic moments \cite{arXiv,EPL}. The
main result can be formulated briefly as follows. For magic
nuclei, the PC corrections are very small, much less than $0.1
\mu_N$, and can not help in solving the problem of $^{209}$Bi. For
semimagic nuclei, they are larger but usually do not exceed $0.1
\mu_N$. However, there are cases of the proton-odd semimagic nuclei
where they reach  $(0.3 \div 0.4) \mu_N$. In particular, the PC
corrections do solve the problem of $^{125,127}$In nuclei mentioned
above. Thus, the one-quasiparticle model works well in magic nuclei
and in the major part of semimagic nuclei but in the last case the
PC corrections are sometimes important.

In view of the overall success of this approach in odd magic and
semimagic nuclei it's seems reasonable to extend it for magnetic
moments of odd-odd nuclei where each odd nucleon  component is odd
magic or only one of them is odd  semimagic. We use the same
strategy supposing in the first step that the odd components satisfy
the single-quasiparticle approximation. In addition, we use the
simplest model of no interaction between the odd particles. Going
beyond this model is rather complicated as one should solve  some
RPA-like equations in the particle-particle  or particle-hole
channels with the corresponding effective interactions which are not
known sufficiently well.  For some nearmagic odd-odd nuclei it has been
made in the works by Isakov {\it et al.}, see for example
\cite{Is1,Is2}, in order to describe the excited states of these
nuclei. In particular,  it was found for $^{134}$Sb  that  the
low-lying states are ''rather pure'' \cite{Is1}. This consideration
confirms our ansatz. In addition, this ansatz can be checked in a
pure phenomenological way for nuclei with known experimental values
of magnetic moments of odd nucleon subsystems. As we will see, this
model does is confirmed generally by a reasonable agreement with the
available experimental data. However, there are several cases where
this simplest model does not work and a configuration mixing
 should be taken into account.

\section{Magnetic moments of odd nearmagic and semimagic nuclei}

The general calculation scheme for odd nuclei is described in detail
in \cite{mu1}. Here we write down only several formulas which are
necessary to understand the  main ingredients of our approach.
Within the TFFS, the static magnetic  moment $\mu_{\lambda}$ of an
odd nucleus with the  odd nucleon in the state
$|\lambda\rangle{=}|n,l,j,m\rangle\equiv|\nu,m\rangle$ can be found
in terms of the diagonal matrix element $ \langle\lambda| V(\omega
=0)|\lambda\rangle$  of the effective field $V$ in the static
external field $V^0$,

\bea \mu _{\tau}&=&\langle \nu,m=j| V ^{\tau}(\omega=0)|\nu,m=j
\rangle
\nonumber\\
&=&S_j^\mu \langle \nu\parallel V ^{\tau}(\omega=0)\parallel
\nu\rangle, \eea where $\tau = p,n$  and
$S_j^\mu$=$\sqrt{\frac{j}{(j+1)(2j+1)}}$ is the corresponding
$3j$-symbol value.

The equation  for the effective
field is written in the representation of the single-particle wave functions  as
 \beq
 \label{Vef_s}
   V_{12}(\omega)=e_{q}V^0_{12}(\omega)+ {\cal F}_{1234}  A_{34}(\omega)  V_{34}(\omega),
\eeq where the short notation $1 \equiv \lambda_1$ is used. For the
static case we consider and nuclei without pairing, the propagator
$A_{12}$ is as follows \beq A_{12} = \frac{n_1 -n_2}{\varepsilon_1
-\varepsilon_2} \eeq where $n_1= (0,1)$ are the occupation numbers.

In the case of nuclei with pairing,  in general it is necessary to
solve   a set of three  equations instead of one Eq. (\ref{Vef_s}),
which contains  two additional equations for the change of the
pairing gap, $d_1$ and $d_2$, in the external field $V_0$, see
\cite{AB1}. However, as it was found in \cite{mu1}, their
contribution for the magnetic moment problem is small, therefore we
will not take them into account. In this case, only the quantity $A$
in Eq. (\ref{Vef_s}) is changed for nuclei  with pairing: \beq
A_{12} = - \frac{E_{1}E_{2} -\varepsilon_1 \varepsilon_2 -\Delta_{1}
\Delta_{2}}{2E_{1}E_{2}(E_{1}+E_{2})} \eeq where $E_{1} =
\sqrt{(\varepsilon_{1}-\mu)^2 + \Delta_{1}^2}$, $\mu$ and
$\Delta_{1}$ being the chemical potential and the diagonal matrix
element of the gap $\Delta$, correspondingly.  The effective
spin-dependent LM interaction is: \beq {\cal F}{=}C_0 \left( g {+}g'
{\bfg \tau}_1 {\bfg \tau}_2 \right) {\bfg \sigma}_1 {\bfg \sigma}_2
\delta({\bf r}_1{-}{\bf r}_2) \delta({\bf r}_2{-}{\bf
r}_3)\delta({\bf r}_3{-}{\bf r}_4).\eeq

After separation of angular parts in Eq.(\ref{Vef_s}) one obtains

\bea \label{main1} \langle \nu_{1}\Vert V \Vert
\nu_{2}\rangle^{\tau} = I^{\tau}_{\nu_{1}\nu_{2}}(e_{qs}^{\tau}
\langle \nu_{1}\Vert T^{1}_{01}\Vert \nu_{2}\rangle^{\tau}  \nonumber\\
+ g_l^{\tau}\delta_{j_{1}j_{2}}\sqrt{j_{1}(j_{1}+1)(2j_{1}+1)})  \nonumber\\
+ \sum_{\tau^{\prime} \nu_3 \nu_4}  F^{\tau
\tau^\prime}_{\nu_{1}\nu_{2}\nu_{3}\nu_{4}}
A^{\tau^\prime}_{\nu_{3}\nu_{4}} \langle \nu_{3}\Vert V \Vert
\nu_{4}\rangle^{\tau^\prime}.
\eea Here

\beq I^{\tau}_{\nu_{1}\nu_{2}} = \int
R^{\tau}_{\nu_{1}}R^{\tau}_{\nu_{2}}r^{2}dr \eeq

and \bea F^{\tau \tau^\prime}_{\nu_{1}\nu_{2}\nu_{3}\nu_{4}} &=&
\frac{C_0}{12\pi}g^{\tau \tau^\prime} \Lambda^{\tau
\tau^\prime}_{\nu_{1}\nu_{2}\nu_{3}\nu_{4}} ( \langle\nu_{1}\Vert
T^{1}_{01}\Vert \nu_{2}\rangle^{\tau}\langle\nu_{1}\Vert
T^{1}_{01}\Vert \nu_{2}\rangle^{\tau^{\prime}}
\nonumber\\
&+& 5 \langle\nu_{1}\Vert T^{1}_{21}\Vert
\nu_{2}\rangle^{\tau}\langle\nu_{1}\Vert T^{1}_{21}\Vert
\nu_{2}\rangle)^{\tau^{\prime}},
\eea

\beq g^{nn}=g^{pp}=g+g',\quad g^{np}=g^{pn}=g-g', \eeq

 \beq  \Lambda^{\tau \tau^\prime}_{\nu_{1}\nu_{2}\nu_{3}\nu_{4}} =
 \int R^{\tau}_{\nu_{1}}R^{\tau}_{\nu_{2}}R^{
\tau^\prime}_{\nu_{3}}R^{\tau^\prime}_{\nu_{4}}r^{2}dr,
\eeq
 where
$R^{\tau}_{\nu_{i}}$ are the  radial single-particle wave functions.

The local quasi-particle charge in Eq.(\ref{Vef_s}) is as follows

\bea e_q{\hat {\bfg{\mu}}}=\frac  {\,{1+(1-2\zeta_{\,l}){\hat
{\tau_3}}}} 2  {\hat  {\textbf{j}}}
\qquad \qquad \qquad \qquad \qquad \qquad \qquad \nonumber\\
+ \frac {  { \left(\gamma_p{+}\gamma_n {-} \frac 1 2 \right) {+}
\left[(\gamma_p{-}\gamma_n)(1{-}2\zeta_s) {-}
 \frac 1 2 {+}\zeta_{\,l} \right] {\hat {\tau_3}}} } 2 \;
 \hat{\bfg {\sigma}},\quad \;
\eea

so that the values $e_{qs}^p, e_{qs}^n, g^p_l, g^n_l$  in
Eq.(\ref{main1}) are \beq e_{qs}^p = (1-\zeta_s)\gamma^p + \zeta_s
\gamma^n - \frac{1}{2} + \frac{1}{2}\zeta_{\,l},\hspace{0.3cm}
 g^p_l = 1-\zeta_{\,l},
 \eeq
 \beq
 e_{qs}^n = (1-\zeta_s)\gamma^n + \zeta_s \gamma^p + \frac{1}{2}\zeta_{\,l},\hspace{0.3cm}
 g^n_{\,l} = \zeta_{\,l}.
\eeq The parameter values $C_0$, $\zeta_{\,l}$, $\zeta_s$, $g,\;g' $
which we used to solve Eq.(\ref{main1}) are as follows: $C_0 =$ 300
MeV fm$^3$, $\zeta_{\,l}$ = -0.05, $\zeta_s$ = 0.08, $g=0.1$,
$g'=1.1$ \cite{mu1}.

\begin{table}[tbp]
\caption{Magnetic moments($\mu_N$) of nearmagic odd nuclei.}
\begin {tabular}{l c c c c}
\hline\noalign{\smallskip}
Nucleus & $\lambda$ &  $\mu_{\rm th}$ \cite{mu1,mu2}  & $\mu_{\rm th}$ & $\mu_{\rm exp}$ \cite{stone,BNL}  \\

\noalign{\smallskip}\hline\noalign{\smallskip}

${^{15}_{7}}$N$_{8}$ & $1p_{1/2}^p$ &  & -0.318 & -0.2831888 (5)  \\
 ${^{15}_{8}}$O$_{7}$& $1p_{1/2}^n$ &  & +0.732 & 0.7189 (8)       \\
 ${^{17}_{8}}$O$_{9}$& $1d_{5/2}^n$ &  & -1.764 & -1.89379 (9)    \\
${^{17}_{9}}$F$_{8}$ & $1d_{5/2}^p$ &  & +4.619 & +4.7223 (12)    \\

\noalign{\smallskip}\hline\noalign{\smallskip}

 ${^{39}_{19}}$K$_{20}$ & $1d_{3/2}^p$  & +0.329 & +0.352 & +0.391466 (33)  \\
 ${^{39}_{20}}$Ca$_{20}$& $1d_{3/2}^n$  & +0.888 & +0.884 & 1.02168 (12)  \\
${^{41}_{20}}$Ca$_{21}$ & $1f_{7/2}^n$ & -1.626 & -1.645 & -1.594781 (9)   \\
 ${^{41}_{21}}$Sc$_{20}$& $1f_{7/2}^p$  & +5.485 & +5.458 & +5.4305 (18)   \\

\noalign{\smallskip}\hline\noalign{\smallskip}

 ${^{55}_{27}}$Co$_{28}$&  $1f_{7/2}^p$  & & +4.793 & 4.822(3)             \\
${^{55}_{28}}$Ni$_{27}$ &  $1f_{7/2}^n$  & & -1.002 & -0.976 (26)              \\
 ${^{57}_{28}}$Ni$_{29}$&  $2p_{3/2}^n$  & -0.882 & -0.812 & -0.7975 (14)  \\
 ${^{57}_{29}}$Cu$_{28}$&  $2p_{3/2}^p$  &  & +2.343 & 2.00 (5)            \\

\noalign{\smallskip}\hline\noalign{\smallskip}

${^{131}_{49}}$In$_{82}$ &  $1g_{9/2}^p$   &  &+5.427 &                    \\
 ${^{131}_{50}}$Sn$_{81}$&  $2d_{3/2}^n$ & +0.714  & +0.724  & 0.747 (4)  \\
 ${^{133}_{50}}$Sn$_{83}$&  $2f_{7/2}^n$  &  & -1.132 &                   \\
${^{133}_{51}}$Sb$_{82}$ &  $1g_{7/2}^p$ & +2.693   & +2.812 & 3.00 (1)   \\
\noalign{\smallskip}\hline\noalign{\smallskip}

 ${^{207}_{81}}$Tl$_{126}$& $3s_{1/2}^p$ &  +1.857 & +1.923 & +1.876 (5)    \\
 ${^{207}_{82}}$Pb$_{125}$& $3p_{1/2}^n$ &  +0.475 & +0.603 & +0.593 (9)   \\
 ${^{209}_{82}}$Pb$_{127}$& $2g_{9/2}^n$ &  -1.335 & -1.452 & -1.4735 (16)  \\
${^{209}_{83}}$Bi$_{126}$ & $1h_{9/2}^p$ &  +3.691 & +3.642 & +4.1106 (2)    \\
 \noalign{\smallskip}\hline
 \end{tabular}
 \end{table}

\begin{table}[tbp]
\caption{Magnetic moments ($\mu_N$) of semimagic
$n$-odd nuclei.
(asterisks note nuclei in  excited state, see sect. 3.2.2)}
\begin {tabular}{l c c c c c}
\hline\noalign{\smallskip}
 Nucleus & $\lambda$&   $   \mu_{\rm th} $ \cite{mu1,mu2}  & $ \mu_{\rm th} $ & $\mu_{\rm exp}$
 \cite{stone,BNL} &\\
\noalign{\smallskip}\hline\noalign{\smallskip}
 ${^{59}_{28}}$Ni$_{31}$&   $2p_{3/2}$ &  &-0.877&&\\

\noalign{\smallskip}
 ${^{59}_{28}}$Ni$_{31}^{\ast}$&   $1f_{5/2}$ & +0.494  &+0.563& +0.35 (15)&\\

\noalign{\smallskip}
 ${^{61}_{28}}$Ni$_{33}$&   $2p_{3/2}$ &  -0.897 &-0.829&-0.75002 (4)&\\

\noalign{\smallskip}
 ${^{87}_{36}}$Kr$_{51}$ &  $2d_{5/2}$& &-1.074&-1.023 (2)&\\

\noalign{\smallskip}
 ${^{89}_{38}}$Sr$_{51}$&  $2d_{5/2}$ & &-1.155&-1.1481 (8)&\\

\noalign{\smallskip}
 ${^{89}_{40}}$Zr$_{49}$&  $1g_{9/2}$ &  &-1.142&-1.076 (20)&\\

\noalign{\smallskip}
 ${^{91}_{40}}$Zr$_{51}$ &  $2d_{5/2}$& &-1.457&-1.30362 (2)&\\

\noalign{\smallskip}
 ${^{91}_{42}}$Mo$_{49}$&  $1g_{9/2}$ & &-1.149&&\\

\noalign{\smallskip}
  ${^{93}_{42}}$Mo$_{51}$&  $2d_{5/2}$ &  &  -1.150 &&\\

\noalign{\smallskip}
 ${^{95}_{44}}$Ru$_{51}$ &  $2d_{5/2}$& &-0.823&-0.861 (7)&\\

\noalign{\smallskip}
 ${^{107}_{50}}$Sn$_{57}$ &  $2d_{5/2}$&&-0.908&&\\

 \noalign{\smallskip}
 ${^{107}_{50}}$Sn$_{57}^{\ast}$ &  $1g_{7/2}$&&+0.714&&\\

\noalign{\smallskip}
 ${^{111}_{50}}$Sn$_{61}$&  $1g_{7/2}$ &  &+0.568&+0.608 (4)&\\

\noalign{\smallskip}
 ${^{111}_{50}}$Sn$_{61}^\ast$&  $2d_{5/2}$ & &-1.046& &\\

\noalign{\smallskip}
 ${^{113}_{50}}$Sn$_{63}$ &  $3s_{1/2}$& -1.025 &-0.977&-0.8791 (6)&\\

\noalign{\smallskip}
 ${^{115}_{50}}$Sn$_{65}$&   $3s_{1/2}$ &-1.030 &-1.019&-0.91883 (7)&\\

\noalign{\smallskip}
 ${^{123}_{50}}$Sn$_{73}$ &  $1h_{11/2}$& -1.307&-1.470&-1.3700 (9)&\\

\noalign{\smallskip}
 ${^{125}_{50}}$Sn$_{75}$ &  $1h_{11/2}$& -1.286&-1.459&-1.348 (2)&\\

\noalign{\smallskip}
 ${^{127}_{50}}$Sn$_{77}$ &  $1h_{11/2}$& -1.268 &-1.456&-1.329 (7)&\\

\noalign{\smallskip}
 ${^{135}_{54}}$Xe$_{81}$ &  $2d_{3/2}$& &+0.818&+0.9032 (7)&\\

\noalign{\smallskip}
 ${^{137}_{54}}$Xe$_{83}$ &  $2f_{7/2}$& &-0.986&-0.9704 (10)&\\

\noalign{\smallskip}
 ${^{137}_{56}}$Ba$_{81}$ &  $2d_{3/2}$& &+0.882&+0.935814 (5)&\\

\noalign{\smallskip}
 ${^{139}_{56}}$Ba$_{83}$ &  $2f_{7/2}$& &-0.985&-0.973 (5)&\\

\noalign{\smallskip}
 ${^{139}_{58}}$Ce$_{81}$ &  $2d_{3/2}$& &+1.120&1.06 (4)&\\

\noalign{\smallskip}
 ${^{141}_{58}}$Ce$_{83}$ &  $2f_{7/2}$& &-0.967&1.09 (4)&\\

\noalign{\smallskip}
 ${^{143}_{60}}$Nd$_{83}$ &  $2f_{7/2}$& &-0.950&-1.065 (5)&\\

\noalign{\smallskip}
 ${^{143}_{62}}$Sm$_{81}$ &  $2d_{3/2}$& &+0.983&+1.01 (2)&\\

\noalign{\smallskip}
 ${^{145}_{62}}$Sm$_{83}$ &  $2f_{7/2}$& &-1.145&-1.11 (6)&\\

\noalign{\smallskip}
 ${^{147}_{64}}$Gd$_{83}$ &  $2f_{7/2}$& &-1.138&1.02 (9)&\\

\noalign{\smallskip}
 ${^{193}_{82}}$Pb$_{111}$&  $3p_{3/2}$ & -1.106 &-0.980&&\\

 \noalign{\smallskip}
 ${^{193}_{82}}$Pb$_{111}^\ast$&  $2f_{5/2}$ & +0.677 &+0.603&&\\

\noalign{\smallskip}
 ${^{195}_{82}}$Pb$_{113}$&  $3p_{3/2}$ &-1.090 &-0.996&&\\

 \noalign{\smallskip}
 ${^{195}_{82}}$Pb$_{113}^\ast$&  $2f_{5/2}$ & +0.672 &+0.680&&\\

\noalign{\smallskip}
 ${^{197}_{82}}$Pb$_{115}$ &  $3p_{3/2}$ & -1.090 &-1.109&-1.075 (2)&\\

 \noalign{\smallskip}
 ${^{197}_{82}}$Pb$_{115}^\ast$ &  $2f_{5/2}$ & +0.668 &+0.728&&\\

\noalign{\smallskip}
 ${^{199}_{82}}$Pb$_{117}$&   $3p_{3/2}$ & -1.090 &-1.101&-1.0742 (12)&\\

 \noalign{\smallskip}
 ${^{199}_{82}}$Pb$_{117}^\ast$&   $2f_{5/2}$ & +0.669 &+0.752&&\\

\noalign{\smallskip}
 ${^{201}_{82}}$Pb$_{119}$&   $2f_{5/2}$ & +0.670 &+0.717&+0.6753 (5)&\\

 \noalign{\smallskip}
 ${^{201}_{82}}$Pb$_{119}^\ast$&   $3p_{3/2}$ &  &-1.002&&\\

\noalign{\smallskip}
 ${^{203}_{82}}$Pb$_{121}$ &  $2f_{5/2}$ & +0.677 &+0.721&+0.6864 (5)&\\

 \noalign{\smallskip}
 ${^{203}_{82}}$Pb$_{121}^\ast$ &  $3p_{3/2}$ &  &-1.103&&\\

\noalign{\smallskip}
 ${^{205}_{82}}$Pb$_{123}$&   $2f_{5/2}$ & +0.690 &+0.666&+0.7117 (4)&\\

 \noalign{\smallskip}
 ${^{205}_{82}}$Pb$_{123}^\ast$&   $3p_{3/2}$ &  &-1.116&&\\

\noalign{\smallskip}
 ${^{211}_{82}}$Pb$_{129}$&   $2g_{9/2}$ & -1.316 &-1.253&-1.4037 (8)&\\

\noalign{\smallskip}
 ${^{211}_{86}}$Rn$_{125}$&   $3p_{1/2}$ & &+0.566&+0.601 (7)&\\

\noalign{\smallskip}
 ${^{213}_{88}}$Ra$_{125}$&  $3p_{1/2}$ & &+0.547&+0.6133 (18)&\\
\noalign{\smallskip}\hline
\end{tabular}
\end{table}

\begin{table}[tbp]
\caption{Magnetic moments ($\mu_N$) of semimagic
$p$-odd nuclei.
(asterisks note nuclei in  excited state, see sect. 3.2.2)}
\begin {tabular}{lc c c c c c  } \hline\noalign{\smallskip}
  Nucleus & $\lambda$&  $\mu_{\rm th} $ \cite{mu1,mu2} & $\mu_{\rm th}$ & $\mu_{\rm exp}$  \cite{stone,BNL}\\
\noalign{\smallskip}\hline\noalign{\smallskip}
 ${^{57}_{27}}$Co$_{30}$ &  $1f_{7/2}$& &+4.899&+4.720 (10)\\

\noalign{\smallskip}
${^{59}_{27}}$Co$_{32}$ &   $1f_{7/2}$&  &+4.609&+4.627 (9)\\

\noalign{\smallskip}
${^{61}_{27}}$Co$_{34}$ &   $1f_{7/2}$& &+4.802&\\
\noalign{\smallskip}
${^{87}_{37}}$Rb$_{50}$&    $2p_{3/2}$ & &+2.663&+2.75131 (12)\\

\noalign{\smallskip}
${^{89}_{39}}$Y$_{50}$&     $2p_{1/2}$ & &-0.153&-0.1374154 (3)\\

\noalign{\smallskip}
${^{91}_{41}}$Nb$_{50}$&    $1g_{9/2}$ & &+6.203&\\

\noalign{\smallskip}
${^{93}_{43}}$Tc$_{50}$&    $1g_{9/2}$ & +6.196 &+6.454&6.32 (6)\\

\noalign{\smallskip}
${^{105}_{49}}$In$_{56}$ &  $1g_{9/2}$& &+5.451&+5.675 (5)\\

\noalign{\smallskip}
${^{107}_{49}}$In$_{58}$ &  $1g_{9/2}$& &+5.312&+5.585 (8)\\

\noalign{\smallskip}
${^{109}_{49}}$In$_{60}$&   $1g_{9/2}$ & &+5.359&+5.538 (4)\\

\noalign{\smallskip}
${^{111}_{49}}$In$_{62}$&   $1g_{9/2}$ & &+5.413&+5.503 (7)\\

\noalign{\smallskip}
${^{113}_{51}}$Sb$_{62}$&   $2d_{5/2}$ & &+3.926&\\

\noalign{\smallskip}
${^{115}_{51}}$Sb$_{64}$ &  $2d_{5/2}$& &+3.526&+3.46 (1)\\

\noalign{\smallskip}
${^{117}_{51}}$Sb$_{66}$ &  $2d_{5/2}$&&+3.378&+3.43 (6)\\

\noalign{\smallskip}
${^{123}_{51}}$Sb$_{72}$ &  $1g_{7/2}$& +2.661 &+2.484&+2.5498 (2)\\

\noalign{\smallskip}
${^{123}_{51}}$Sb$_{72}^\ast$ &  $2d_{5/2}$&  &+3.472&\\

\noalign{\smallskip}
${^{125}_{51}}$Sb$_{74}$  &  $1g_{7/2}$& +2.671 &+2.724&+2.63 (4)\\

\noalign{\smallskip}
${^{125}_{51}}$Sb$_{74}^\ast$  &  $2d_{5/2}$&  &+3.472&\\

\noalign{\smallskip}
${^{127}_{51}}$Sb$_{76}$&    $1g_{7/2}$ & +2.682 &+2.756&2.697 (6)\\

\noalign{\smallskip}
${^{129}_{51}}$Sb$_{78}$ &  $1g_{7/2}$& +2.692 &+2.793&2.817 (5)\\

\noalign{\smallskip}
${^{137}_{55}}$Cs$_{82}$ &  $1g_{7/2}$& +2.577 &+2.682&+2.8413 (1)\\

\noalign{\smallskip}
${^{139}_{57}}$La$_{82}$&   $1g_{7/2}$ & +2.545 &+2.822&+2.7830455 (9)\\

\noalign{\smallskip}
 ${^{141}_{59}}$Pr$_{82}$ &  $2d_{5/2}$&+4.122 &+4.159&4.2754 (5)\\

\noalign{\smallskip}
${^{145}_{63}}$Eu$_{82}$ &   $2d_{5/2}$&+4.122 &+4.126&+3.993 (7)\\

\noalign{\smallskip}
${^{191}_{81}}$Tl$_{110}$ &  $3s_{1/2}$& &+1.846&1.588 (4)\\

\noalign{\smallskip}
${^{193}_{81}}$Tl$_{112}$ &  $3s_{1/2}$& &+1.680&+1.5912 (22)\\

\noalign{\smallskip}
${^{195}_{81}}$Tl$_{114}$ &  $3s_{1/2}$& &+1.579&+1.58 (4)\\

\noalign{\smallskip}
${^{197}_{81}}$Tl$_{116}$ &  $3s_{1/2}$& &+1.492&+1.58 (2)\\

\noalign{\smallskip}
${^{199}_{81}}$Tl$_{118}$&   $3s_{1/2}$ & &+1.506&+1.60 (2)\\

\noalign{\smallskip}
${^{201}_{81}}$Tl$_{120}$ &  $3s_{1/2}$& &+1.720&+1.605 (2)\\

\noalign{\smallskip}
${^{203}_{81}}$Tl$_{122}$ &  $3s_{1/2}$&&+1.663&+1.62225787 (12)\\

\noalign{\smallskip}
${^{205}_{81}}$Tl$_{124}$&   $3s_{1/2}$ &&+1.951&+1.63821461 (12)\\

\noalign{\smallskip}
${^{201}_{83}}$Bi$_{118}$ &   $1h_{9/2}$& &+4.670&4.8 (3)\\

\noalign{\smallskip}
${^{203}_{83}}$Bi$_{120}$ &   $1h_{9/2}$& &+3.861&+4.017 (13)\\

\noalign{\smallskip}
${^{205}_{83}}$Bi$_{122}$ &   $1h_{9/2}$& &+3.757&+4.065 (7)\\

\noalign{\smallskip}
${^{207}_{83}}$Bi$_{124}$&    $1h_{9/2}$ & &+3.868&+4.0915 (9)\\

\noalign{\smallskip}
${^{211}_{83}}$Bi$_{128}$&    $1h_{9/2}$ & &+3.475&3.5 (3)\\

\noalign{\smallskip}
${^{213}_{83}}$Bi$_{130}$ &   $1h_{9/2}$& &+3.618&+3.717 (13)\\

\noalign{\smallskip}
${^{213}_{87}}$Fr$_{126}$ &   $1h_{9/2}$ & +3.480 &+3.980&+4.02 (8)\\

\hline
\end{tabular}
\end{table}

 Eq. (\ref{main1})
was  solved in the  representation of the self-consistent
 single-particle wave functions
on the base of the functional by Fayans {\it et al.} \cite{Fay} with the set
 DF3a \cite{Tol-Sap} of the parameters.
To reliably take into account the single-particle continuum we used
the spherical box with the radius $R = 24$ cm and the upper cut for
the single-particle spectrum equal to 50 MeV. In order to check the
convergence, the upper cut was taken as 100 MeV for several isotopes
of Sn, Pb and Bi. It turned out that the difference from the 50 MeV
cut was less than 1\%. Like in \cite{mu1,Q}, to obtain the magnetic
moments values in semimagic odd nuclei, we used the half-sum of the
magnetic moments values (which, as a rule, are very
close to each other) in two
neighboring even nuclei.

Thus, the present calculations differ from those  of Ref. \cite{mu1}
in the following: i) the representation of  single-particle wave
functions has been used, ii)we have used a slightly simplified
version of the effective forces as compared with Ref. \cite{mu1}, in
particular, the $\pi$- and $\rho$-exchange forces  have not been
considered. Together with our main approximation, i.e. the neglect
of the interaction between odd proton and odd neutron, all our
approximations  probably have the  similar order of magnitude.
It is clear, however, that our simple model to calculate magnetic
moments of odd-odd nuclei within such an approximation   should be
checked quantitatively before  the above-mentioned corrections are
included, and this check
 should be done for all of them in order to be within the same level of accuracy. For the purpose
 of comparison, the results   of Ref. \cite{mu1} are  given  in Tables 1, 2 and 3
for nuclei common for the present calculation and that of Ref. \cite{mu1}, it
is  about 30\% of the total number of odd nuclei considered here. We  see that
the agreement between  our results and the results of Ref.
\cite{mu1} is rather good. In addition to the above arguments,  we
should note that the agreement with the experiment obtained in
 Ref. \cite{mu1} is  within 0.1--0.2 $\mu_N$, which is almost always  the same for our results, and  this fact
 justifies  partly our approximations.

We calculated  only magnetic moments of those odd nuclei which
are  necessary to construct the odd-odd ones under consideration,
see below  our selection for them.

The results of the calculations are presented in Table 1 for
magic odd nuclei around double magic $^{16}$O, $^{40}$Ca,
$^{56}$Ni, $^{132}$Sn and $^{208}$Pb and in Tables 2 and 3 for
odd semimagic nuclei.
For nearmagic odd nuclei, Table 1, we obtained a good agreement with the experiment, except for $^{209}$Bi,
see discussion in Introducion.
As for semimagic odd nuclei, a good agreement with the experiment is also obtained  with  the same accuracy
0.1--0.2 $\mu_N$ as in Ref. \cite{mu1}, except for three cases where the value $|\mu_{\rm th} - \mu_{\rm exp}|$ is
0.26 $\mu_N$ and 0.31 $\mu_N$  for $^{191}$Tl and $^{205}$Bi, respectively.

\section{Magnetic moments of odd-odd nuclei}

If we neglect the interaction between two odd quasi-particles,
the magnetic  moment of an odd-odd nucleus with the spin $J$ is as
follows:

\beq \label{qu} \mu_J=\left(\Psi_{JJ} [V^p + V^n] \Psi_{JJ}\right),
\eeq where $\Psi_{JJ} = \Sigma \varphi_1 \varphi_2 <j_1m_1
j_2m_2\mid JJ>$
and $\varphi_1$ is the
single-particle wave function with the quantum numbers $1 \equiv
\lambda_1 \equiv (n_1,j_1,l_1,m_1)$.
Then the expression for the ground state magnetic moment of the odd-odd
nucleus is as follows:
\bea
\label{main}
\mu=\frac{J \mu_p}{ 2j_p} \left( 1+\frac{(j_p-j_n)(j_p+j_n+1)}{J(J+1)} \right)+ \nonumber\\
 \frac{J \mu_n}{ 2j_n} \left(1+\frac{(j_n-j_p)(j_p+j_n+1)}{J(J+1)} \right)
\eea
where $\mu_n$ and $\mu_p$ are taken from Eq.(1). Eq.(\ref{main}) is also true for
three other  cases with odd holes.

Thus, within such a simple approximation, the problem is reduced to
the calculations of magnetic moments of corresponding odd-even
nuclei. One can take the values of magnetic moments of odd nuclei
from the experiment (phenomenological approach ),  and   one can
also calculate them microscopically with Eq. (\ref{main1}) and
obtain the magnetic moments of corresponding odd-odd nuclei
according to Eq. (\ref{main}).
We see that there is  a pure phenomenological method to check our
main approximation to neglect  the interaction between two odd
quasi-particles, if there are experimental data on magnetic moments
of two odd corresponding nuclei. It turned out that there are a lot
of  cases like these. In all calculations for odd-odd nuclei given
in Tables 4-8 we show the results of our phenomenological
calculations (column $\mu_{\rm phen}$) made  with the use of
experimental values of ground state magnetic moments of odd nuclei,
according to Eq. (\ref{main}). A very reasonable general agreement
with the experiment for magic nuclei as well as for semimagic
ones confirms our main approximation of no
 interaction between odd quasi-particles.


\begin{table*}[tbp]
\caption{Magnetic moments  ($\mu_N$) of  odd-odd nearmagic nuclei.}
\begin {tabular}{l c c c c c c c c}
\hline
  Nucleus & $\lambda_1^n\lambda_2^p$& $T_{\rm 1/2}$ & $J^{\rm \pi}$ & $   \mu_{\rm Sch} $ &
  $\mu_{\rm phen}$ & $\mu_{\rm th}$ & $\mu_{\rm exp} $\cite{stone,BNL}  \\
\hline
${^{14}_{7}}$N$_{7}$&(1p$_{1/2}^n$ 1p$_{1/2}^p$) & Stable & $1^+$ & +0.374  & +0.4370 (8)& +0.414 & +0.40376100 (6)\\
 ${^{16}_{7}}$N$_{9}$&(1d$_{5/2}^n$ 1p$_{1/2}^p$) & 7.13 s & $2^-$ & +1.961  & +1.95623 (9)& +1.858  & 1.9859 (11) \\
${^{18}_{9}}$F$_{9}$&(1d$_{5/2}^n$ 1d$_{5/2}^p$) & 109.77 min & $1^+$ & +0.576 & +0.5654 (12) &+0.571 & -\\
\hline
 ${^{38}_{19}}$K$_{19}$ &(1d$_{3/2}^n$ 1d$_{3/2}^p$)& 7.636 min & $3^+$ & +1.272  & +1.41315 (12)& +1.236 & +1.371 (6)\\
 ${^{40}_{19}}$K$_{21}$ &(1f$_{7/2}^n$ 1d$_{3/2}^p$)& 1.246*10$^9$ y & $4^-$ & -1.683  & -1.249267 (20) & -1.316 & -1.2981 (3) \\
${^{40}_{21}}$Sc$_{19}$ &(1d$_{3/2}^n$ 1f$_{7/2}^p$)& $182.3 ms$ & $4^-$ & +5.908  & +5.5114 (16)& +5.462 & - \\
\hline
${^{56}_{27}}$Co$_{29}$ & (1f$_{7/2}^n$ 2p$_{3/2}^p$)& 77.236 d  & $4^+$ & +4.276  & +3.983 (3) & +3.949& 3.85 (1) \\
 ${^{56}_{29}}$Cu$_{27}$ & (2p$_{3/2}^n$ 1f$_{7/2}^p$)& 93 ms & $(4^+)$ & +0.274  &+0.17 (5)& +0.333  & - \\
 ${^{58}_{29}}$Cu$_{29}$ &(2p$_{3/2}^n$ 2p$_{3/2}^p$)& 3.204 s & $1^+$ & +0.627  & +0.47 (5) & +0.510 & 0.52 (8) \\
\hline
 ${^{132}_{49}}$In$_{83}$ &(2d$_{3/2}^n$ 1g$_{7/2}^p$)& 0.207 s & $(7^-)$ & +4.527   & -& +3.949 & - \\
 ${^{132}_{51}}$Sb$_{81}$ &(2f$_{7/2}^n$ 1g$_{9/2}^p$)& 2.79 min & $(4^+)$ & +2.274  & +3.141 (7)& +2.957 & 3.18 (1)  \\
\hline
 ${^{208}_{81}}$Tl$_{127}$ &(2g$_{9/2}^n$ 3s$_{1/2}^p$)& 3.053 min & $5^+$ & +0.880  & +0.402 (5)& +0.471 & 0.292 (13) \\
 ${^{208}_{83}}$Bi$_{127}$ &(3p$_{1/2}^n$ 1h$_{9/2}^p$)& 3.68*10$^5$ y & $5^+$ & +3.262  & +4.703 (9)& +4.235 & +4.578 (13) \\
${^{210}_{83}}$Bi$_{129}$ &(2g$_{9/2}^n$ 1h$_{9/2}^p$)& 5.012 d & $1^-$ & +0.078  & +0.2930 (16)& +0.243 & -0.04451 (6) \\
 \hline
\end{tabular}
\end{table*}

In the present article, for the simplicity and definiteness, we
restrict ourselves by using only \textit{ground state} magnetic
moments of odd nuclei in order  to obtain magnetic moments of
odd-odd nuclei, according to Eq. (\ref{main}). This fact restrict
noticeably the selection of  odd-odd nuclei simply because of the
spin and parity selection rules.
However, to describe the magnetic
moments of  odd-odd nuclei for the case of semi-degenerated single-particle levels we
are forced to use the magnetic moments values  of odd nuclei in excited states as well,
see Sec. 3.2.2 and Tables 2 and 3.
Also, we avoided nuclei with presumably strong many-particle correlation effects.

It should be noted that signs of the experimental magnetic moments
values, which are taken from \cite{stone,BNL},
are  not always measured.
In such  cases we predict the  unknown signs.
As far as in all cases of the known  experimental sign, the theoretical one, from both
the columns $\mu_{\rm th}$ and $\mu_{\rm phen}$, is correct, our predictions for
unknown signs should be reliable.

\subsection{Magnetic moments of odd-odd nearmagic nuclei}
The results of the calculations for magic odd-odd nuclei are
given in Table 4. A reasonable agreement with the available experimental
data  was obtained for both the ``theoretical'' columns, $\mu_{\rm th}$  and $\mu_{\rm phen}$,
except for $^{208}$Bi and  $^{210}$Bi isotopes. In the first case, the phenomenological value
agrees with the data reasonably well, hence the discrepancy  of 0.34 $\mu_N$ for $\mu_{\rm th}$ value
can be mainly attributed to the known uncertainty
of the theoretical magnetic moment of $^{209}$Bi as
it was discussed above. In the $^{210}$Bi case, the situation is different. Here both the $\mu_{\rm phen}$
and $\mu_{\rm th}$ values disagree with the data at approximately 0.3 $\mu_N$ hence the correct
inputs of odd components do not save. Evidently, some configuration mixing should be taken into
account. This point should be analyzed separately.
For all other cases, our
main approximation of no interaction between odd quasi-particles is
confirmed.

Taking into account such a reasonable description of the ground
state magnetic moments of nearmagic odd-odd nuclei presented in
Table 4 and the fact that the phonon contribution in
magic nuclei should be small \cite{arXiv,EPL}, we have calculated the magnetic moments in
\textit{excited} states for these nuclei  using, as earlier in Table
4,  the \textit{ground } state phenomenological (if   any) or
theoretical values for magnetic moments of odd nearmagic nuclei.
These results are given in Tables 5 and 6. Unfortunately, the number
of experimental data is scarce but in the five cases where there are
experimental data  our description is reasonable. We
hope that our predictions are reliable enough.


\begin{table*}[ht!]
\caption{Magnetic moments ($\mu_N$) of  odd-odd nearmagic nuclei  in excited states. }
\begin {tabular}{l c c c c c c c c}
\hline
 Nucleus &  $\lambda_1\lambda_2$ & $J^{\rm \pi}$&$T_{\rm 1/2}$  & Level, keV&$\mu_{\rm phen}$ & $\mu_{\rm th}$ & $\mu_{\rm exp}$ \cite{stone,BNL} \\
\hline
${^{16}_{7}}$N$_{9}$ &(1d$_{5/2}^n$ 1p$_{1/2}^p$)& $3^-$  &91.3 ps&298.22 &-2.177&-2.082&\\
\hline
${^{16}_{9}}$F$_{7}$&(1p$_{1/2}^n$ 1d$_{5/2}^p$) & $2^-$ &&424&+3.928&+3.823&\\
${^{16}_{9}}$F$_{7}$ &(1p$_{1/2}^n$ 1d$_{5/2}^p$)& $3^-$ &&721&+5.441&+5.351&\\
\hline
${^{18}_{9}}$F$_{9}$ &(1d$_{5/2}^n$ 1d$_{5/2}^p$)& $3^+$  &46.9 ps&937.20&+1.696&+1.713&1.68(15)\\
${^{18}_{9}}$F$_{9}$ &(1d$_{5/2}^n$ 1d$_{5/2}^p$)& $5^+$ &162 ns&1121.36&+2.827&+2.855&\\
${^{18}_{9}}$F$_{9}$ &(1d$_{5/2}^n$ 1d$_{5/2}^p$)& $1^+$ &&&+0.565&+0.571&\\
${^{18}_{9}}$F$_{9}$ &(1d$_{5/2}^n$ 1d$_{5/2}^p$)& $2^+$ &&&+1.131&+1.142&\\
${^{18}_{9}}$F$_{9}$ &(1d$_{5/2}^n$ 1d$_{5/2}^p$)& $4^+$ &&&+2.616&+2.284&\\
\hline
${^{38}_{19}}$K$_{19}$&(1d$_{3/2}^n$ 1d$_{3/2}^p$) & $1^+$ &7.0 ps&458.46&+0.471&+0.412&\\
${^{38}_{19}}$K$_{19}$&(1d$_{3/2}^n$ 1d$_{3/2}^p$) & $2^+$ &56 fs&2401.07&+0.942&+0.824&\\
\hline
${^{40}_{19}}$K$_{21}$ &(1f$_{7/2}^n$ 1d$_{3/2}^p$)& $3^-$&4.25 ns&29.8299&-1.367&-1.410&-1.29(9)\\
${^{40}_{19}}$K$_{21}$ &(1f$_{7/2}^n$ 1d$_{3/2}^p$)& $2^-$ &0.28 ps&800.1427&-1.628&-1.645&\\
${^{40}_{19}}$K$_{21}$ &(1f$_{7/2}^n$ 1d$_{3/2}^p$)& $5^-$ &0.87 ps&891.398&-1.203&-1.293&\\
\hline
${^{40}_{21}}$Sc$_{19}$ &(1d$_{3/2}^n$ 1f$_{7/2}^p$)& $3^-$ &&34.3&+4.655&+4.678&\\
${^{40}_{21}}$Sc$_{19}$ &(1d$_{3/2}^n$ 1f$_{7/2}^p$)& $2^-$ &&772.1&+3.974&+4.089&\\
${^{40}_{21}}$Sc$_{19}$ &(1d$_{3/2}^n$ 1f$_{7/2}^p$)& $5^-$ &&893.5&+6.452&+6.432&\\
\hline
${^{42}_{21}}$Sc$_{21}$ & (1f$_{7/2}^n$ 1f$_{7/2}^p$)& $1^+$ &28 fs&611.051 &+0.548&+0.545&\\
${^{42}_{21}}$Sc$_{21}$ & (1f$_{7/2}^n$ 1f$_{7/2}^p$)& $3^+$ &31 ps&1490.43 &+1.644&+1.635&\\
${^{42}_{21}}$Sc$_{21}$ & (1f$_{7/2}^n$ 1f$_{7/2}^p$)& $5^+$ &45 ps&1510.10 &+2.740&+2.725&\\
${^{42}_{21}}$Sc$_{21}$ & (1f$_{7/2}^n$ 1f$_{7/2}^p$)& $2^+$ &69 fs&1586.31 &+1.096&+1.090&\\
${^{42}_{21}}$Sc$_{21}$ & (1f$_{7/2}^n$ 1f$_{7/2}^p$)& $4^+$ && &+2.192&+2.180&\\
${^{42}_{21}}$Sc$_{21}$ & (1f$_{7/2}^n$ 1f$_{7/2}^p$)& $6^+$ && &+3.288&+3.270&\\
${^{42}_{21}}$Sc$_{21}$ & (1f$_{7/2}^n$ 1f$_{7/2}^p$)& $7^+$ && &+3.836&+3.815&\\
\hline
${^{54}_{27}}$Co$_{27}$ &(1f$_{7/2}^n$ 1f$_{7/2}^p$)& $1^+$ &&936.90&+0.558&+0.542&\\
${^{54}_{27}}$Co$_{27}$ &(1f$_{7/2}^n$ 1f$_{7/2}^p$)& $2^+$ &&1445.66&+1.116&+1.084&\\
${^{54}_{27}}$Co$_{27}$ &(1f$_{7/2}^n$ 1f$_{7/2}^p$)& $7^+$ &1.48 min&197.0 &+3.906&+3.794&\\
${^{54}_{27}}$Co$_{27}$ &(1f$_{7/2}^n$ 1f$_{7/2}^p$)& $3^+$ && &+1.674&+1.626&\\
${^{54}_{27}}$Co$_{27}$ &(1f$_{7/2}^n$ 1f$_{7/2}^p$)& $4^+$ && &+2.232&+2.168&\\
${^{54}_{27}}$Co$_{27}$ &(1f$_{7/2}^n$ 1f$_{7/2}^p$)& $5^+$ &&&+2.790&+2.710&\\
${^{54}_{27}}$Co$_{27}$ &(1f$_{7/2}^n$ 1f$_{7/2}^p$)& $6^+$ &&&+3.348&+3.252&\\
\hline
${^{56}_{27}}$Co$_{29}$ & (1f$_{7/2}^n$ 2p$_{3/2}^p$)& $3^+$ &$<$ 0.1 ns&158.38 &+4.133&+4.108&\\
${^{56}_{27}}$Co$_{29}$ & (1f$_{7/2}^n$ 2p$_{3/2}^p$)& $5^+$ &0.28 ps&576.50 &+4.025&+3.981&\\
${^{56}_{27}}$Co$_{29}$ & (1f$_{7/2}^n$ 2p$_{3/2}^p$)& $2^+$ &&&+4.665&+4.650&\\
\hline
${^{56}_{29}}$Cu$_{27}$ & (2p$_{3/2}^n$ 1f$_{7/2}^p$)& $2^+$ &&&-2.171&-2.421&\\
${^{56}_{29}}$Cu$_{27}$ & (2p$_{3/2}^n$ 1f$_{7/2}^p$)& $3^+$ && &-0.837&-0.859&\\
${^{56}_{29}}$Cu$_{27}$ & (2p$_{3/2}^n$ 1f$_{7/2}^p$)& $5^+$ &&&+1.023&+1.341&\\
\hline
${^{58}_{29}}$Cu$_{29}$ &(2p$_{3/2}^n$ 2p$_{3/2}^p$)& $3^+$ &0.32 ns&443.64 &+1.419&+1.530&\\
${^{58}_{29}}$Cu$_{29}$ &(2p$_{3/2}^n$ 2p$_{3/2}^p$)& $2^+$ &$>$ 0.66 ps& 1427.85 &+0.946&+1.020&\\
\hline

\end{tabular}
\end{table*}

\begin{table*}[tbp]
\caption{Magnetic moments ($\mu_N$) of  odd-odd nearmagic nuclei  in excited states (continuations).}

\begin {tabular}{l c c c c c c c c }
\hline
 Nucleus & $\lambda_1\lambda_2$& $J^{\rm \pi}$ &$T_{\rm 1/2}$&Level, kev& $\mu_{\rm phen}$ & $\mu_{\rm th}$ & $\mu_{\rm exp} $ \cite{stone,BNL} \\
\hline
${^{132}_{49}}$In$_{83}$ & (2d$_{3/2}^n$ 1g$_{7/2}^p$)& $1^-$ &&&&+3.882&\\
${^{132}_{49}}$In$_{83}$ & (2d$_{3/2}^n$ 1g$_{7/2}^p$)& $2^-$ &&&&+3.177&\\
${^{132}_{49}}$In$_{83}$ & (2d$_{3/2}^n$ 1g$_{7/2}^p$)& $3^-$ &&&&+3.044&\\
${^{132}_{49}}$In$_{83}$ & (2d$_{3/2}^n$ 1g$_{7/2}^p$)& $4^-$ &&&&+3.142&\\
${^{132}_{49}}$In$_{83}$ & (2d$_{3/2}^n$ 1g$_{7/2}^p$)& $5^-$ &&&&+3.353&\\
${^{132}_{49}}$In$_{83}$ & (2d$_{3/2}^n$ 1g$_{7/2}^p$)& $6^-$ &&&&+3.631&\\
${^{132}_{49}}$In$_{83}$ & (2d$_{3/2}^n$ 1g$_{7/2}^p$)& $8^-$ &&&&+4.295&\\
\hline
${^{132}_{51}}$Sb$_{81}$ &(2f$_{7/2}^n$ 1g$_{9/2}^p$)& $3^+$ &15.62 ns&85.55 &+2.571&+1.982&\\
${^{132}_{51}}$Sb$_{81}$ &(2f$_{7/2}^n$ 1g$_{9/2}^p$)& $5^+$ &&162.8 &+3.747&+3.536&\\
${^{132}_{51}}$Sb$_{81}$ &(2f$_{7/2}^n$ 1g$_{9/2}^p$)& $2^+$ &&&+2.073&+1.982&\\
\hline
${^{134}_{51}}$Sb$_{83}$ &(2f$_{7/2}^n$ 1g$_{7/2}^p$)& $1^-$ &&13.0  &&+0.243&\\
${^{134}_{51}}$Sb$_{83}$ &(2f$_{7/2}^n$ 1g$_{7/2}^p$)& $7^-$ &&279  &&+1.682&\\
${^{134}_{51}}$Sb$_{83}$ &(2f$_{7/2}^n$ 1g$_{7/2}^p$)& $2^-$ &&331.1  &&+0.476&\\
${^{134}_{51}}$Sb$_{83}$ &(2f$_{7/2}^n$ 1g$_{7/2}^p$)& $3^-$ &&384.0  &&+0.724&\\
${^{134}_{51}}$Sb$_{83}$ &(2f$_{7/2}^n$ 1g$_{7/2}^p$)& $5^-$ &&441  &&+1.180&\\
${^{134}_{51}}$Sb$_{83}$ &(2f$_{7/2}^n$ 1g$_{7/2}^p$)& $4^-$ &&555.0  &&+0.959&\\
${^{134}_{51}}$Sb$_{83}$ &(2f$_{7/2}^n$ 1g$_{7/2}^p$)& $6^-$ &&617  &&+1.442&\\
\hline
${^{206}_{81}}$Tl$_{125}$ &(3p$_{1/2}^n$ 3s$_{1/2}^p$)& $1^-$ &4.2 ps&304.90 &+2.469&+2.526&\\
\hline
${^{208}_{81}}$Tl$_{127}$ &(2g$_{9/2}^n$ 3s$_{1/2}^p$)& $4^+$ &6.5 ps&39.858 &-2.942&-2.958&\\
\hline
${^{208}_{83}}$Bi$_{125}$ &(3p$_{1/2}^n$ 1h$_{9/2}^p$)& $4^+$ &&63.02  &+3.545&+3.079&\\
\hline
${^{210}_{83}}$Bi$_{127}$ &(2g$_{9/2}^n$ 1h$_{9/2}^p$)& $9^-$ &3.04 x 10$^6$ y&271.31
&+2.637&+2.187&2.728(42)\\
${^{210}_{83}}$Bi$_{127}$ &(2g$_{9/2}^n$ 1h$_{9/2}^p$)& $2^-$ &5.2 ps&319.74  &+0.586&+0.486&\\
${^{210}_{83}}$Bi$_{127}$ &(2g$_{9/2}^n$ 1h$_{9/2}^p$)& $3^-$ &&347.93 &+0.879&+0.729&\\
${^{210}_{83}}$Bi$_{127}$ &(2g$_{9/2}^n$ 1h$_{9/2}^p$)& $7^-$ &57.5 ns&433.49  &+2.051&+1.701&+2.114(49)\\
${^{210}_{83}}$Bi$_{127}$ &(2g$_{9/2}^n$ 1h$_{9/2}^p$)& $5^-$ &37.7 ns&439.20  &+1.465&+1.215&+1.530(45)\\
${^{210}_{83}}$Bi$_{127}$ &(2g$_{9/2}^n$ 1h$_{9/2}^p$)& $4^-$ &$<$ 1.4 ns&502.81 &+1.172&+0.972&\\
${^{210}_{83}}$Bi$_{127}$ &(2g$_{9/2}^n$ 1h$_{9/2}^p$)& $6^-$ &$<$ 1.4 ns&550.00 &+1.758&+1.458&\\
${^{210}_{83}}$Bi$_{127}$ &(2g$_{9/2}^n$ 1h$_{9/2}^p$)& $8^-$ &$<$ 1.7 ps&582.53 &+2.344&+1.944&\\
\hline
\end{tabular}
\end{table*}

\subsection{Magnetic moments of odd-odd  semimagic nuclei}
The results of the calculations of magnetic moments in odd-odd
semimagic nuclei are given in Table 7. In general, the agreement with the
experiment is a little worse, than that for magic nuclei.
As it was discussed above, to estimate accuracy of the model used,
the comparison of the experimental values $\mu_{\rm exp} $  with the phenomenological
ones, $\mu_{\rm phen}$, is of primary importance. There are two cases, $^{94}$Tc and $^{126}$Sb,
with disagreement of about 0.5 $\mu_N$. They need an additional analysis. In the major
part of nuclei considered $\mu_{\rm phen}$ and $\mu_{\rm th}$ agree very well with each other.
There is the only case, when this is not true, the $^{206}$Bi nucleus,
which is caused by the input from the odd $^{207}$Bi isotope, for here
the disagreement is almost the same as for $^{209}$Bi, see Introduction.
 The $^{90}$Nb case should be discussed
separately. Here the experimental magnetic moment of the odd $^{90}$Nb nucleus is not known
and therefore the $\mu_{\rm phen}$ value is absent. In this case, the theoretical prediction
disagrees with the data at approximately 0.5 $\mu_N$ just as for two nuclei discussed above.
In this case, it is not clear either our model
or the input value of the proton magnetic
moment or both are to be blamed for that. As the analysis in \cite{arXiv,EPL} showed,
the phonon corrections to
magnetic moments of the odd-proton nuclei in this region is approximately 0.3 $\mu_N$.
Hence these corrections should be included into the analysis.

Another possible reason for disagreement with the experiment
may be a close position of the single-particle levels in any of two
odd subsystems leading to a strong configuration
mixture. In nuclei with pairing we deal in this subsection there is an additional reason for such closeness
if to compare with magic ones.  Indeed, the single-particle spectrum is now given with the
Bogolyubov energies, $E_{\lambda}=\sqrt{\Delta_{\lambda}^2 + (\eps_{\lambda}-\mu)^2}$.
If we deal with two single-particle states $\lambda_1, \lambda_2$ close to the Fermi level,
$\eps_{\lambda_1}-\mu < \Delta_{\lambda_1}$ and $\eps_{\lambda_2}-\mu < \Delta_{\lambda_2}$,
the energy difference $|E_{\lambda_1} - E_{\lambda_2}|$ is less, and could
be significantly less, than the initial one, $|\eps_{\lambda_1} - \eps_{\lambda_2}|$.
This is the reason why the configuration mixing in semimagic odd-odd nuclei is,
in general,  more important than in the magic ones.

We have found four  nuclei ${^{58}_{27}}$Co$_{31}$ , ${^{106}_{49}}$In$_{57}$
${^{110}_{49}}$In$_{61}$ and ${^{124}_{51}}$Sb$_{73}$ which have
very close  single-particle levels
in one of the odd subsystems. We will consider them together with
 a special case of  odd-odd Tl isotopes in Section 3.2.2 below.

\begin{table*}[tbp]
\caption{Ground state magnetic moments ($\mu_N$)  of  odd-odd semimagic nuclei. }
\begin {tabular}{l c c c c c c }
\hline
   Nucleus & $\lambda_1^n\lambda_2^p$& $J^{\rm \pi}$&$T_{\rm 1/2}$& $\mu_{\rm phen}$ & $\mu_{\rm th}$ &
$\mu_{\rm exp}$ \cite{stone,BNL} \\
\hline

${^{60}_{27}}$Co$_{33}$&(2p$_{3/2}^n$ 1f$_{7/2}^p$)& $5^+$&1925.28 d &+3.877&+3.993&+3.799 (8)\\



${^{90}_{41}}$Nb$_{49}$& (1g$_{9/2}^n$ 1g$_{9/2}^p$)& $8^+$&14.60 h&&+4.404&4.961 (4) \\

${^{94}_{43}}$Tc$_{51}$& (2d$_{5/2}^n$ 1g$_{9/2}^p$)& $7^+$ &293 min&+5.459&+5.468&5.08 (8)\\






${^{126}_{51}}$Sb$_{75}$  & (1h$_{11/2}^n$ 1g$_{7/2}^p$)& $8^-$ &12.35 d&+0.918&+0.907&1.28 (7)\\

${^{128}_{51}}$Sb$_{77}$& (1h$_{11/2}^n$ 1g$_{7/2}^p$) & $8^-$ &9.01 h&+1.090&+0.994&1.31 (19)\\

${^{136}_{55}}$Cs$_{81}$& (2d$_{3/2}^n$ 1g$_{7/2}^p$)& $5^+$ &13.16 d&+3.761&+3.532&+3.711 (15)\\

${^{138}_{55}}$Cs$_{83}$ & (2f$_{7/2}^n$ 1g$_{7/2}^p$)& $3^-$ &33.41 min &+0.802&+0.727&+0.700 (4)\\

${^{138}_{57}}$La$_{81}$ & (2d$_{3/2}^n$ 1g$_{7/2}^p$)& $5^+$ &1.02 * 10$^{11}$ y&+3.781&+3.823&+3.73646 (7)\\

${^{140}_{57}}$La$_{83}$&(2f$_{7/2}^n$ 1g$_{7/2}^p$)& $3^-$ &1.67855 d&+0.751&+0.791&+0.730 (15)\\

${^{142}_{59}}$Pr$_{83}$&(2f$_{7/2}^n$ 2d$_{5/2}^p$)& $2^-$ &19.12 h&+0.382&+0.316&+0.234 (1)\\

${^{144}_{63}}$Eu$_{81}$ & (2d$_{3/2}^n$ 2d$_{5/2}^p$)& $1^+$ &10.2 s&+2.032&+2.101&+1.893 (13)\\

${^{146}_{63}}$Eu$_{83}$& (2f$_{7/2}^n$ 2d$_{5/2}^p$)& $4^-$ &4.61 d&+1.255&+1.265&+1.421 (8)\\

${^{202}_{83}}$Bi$_{119}$ & (2f$_{5/2}^n$ 1h$_{9/2}^p$)& $5^+$ &1.71 h&+4.071&+4.313&+4.259 (14)\\

${^{204}_{83}}$Bi$_{121}$  & (2f$_{5/2}^n$ 1h$_{9/2}^p$)& $6^+$&11.22 h&+4.230&+4.084&+4.322 (15)\\

${^{206}_{83}}$Bi$_{123}$ &  (2f$_{5/2}^n$ 1h$_{9/2}^p$)& $6^+$ &6.243 d&+4.283&+3.957&+4.361 (8)\\

${^{212}_{83}}$Bi$_{129}$ &  (2g$_{9/2}^n$ 1h$_{9/2}^p$) & $1^-$&60.55 min&+0.557&+0.497&0.41 (5)\\

${^{212}_{87}}$Fr$_{125}$& (3p$_{1/2}^n$ h$_{9/2}^p$)& $5^+$ &20.0 min&+4.627&+4.537&4.62 (9)\\
 \hline
\end{tabular}
\end{table*}

\subsubsection{Magnetic moments of odd-odd semimagic nuclei in excited states.}

We have found five
odd-odd semimagic nuclei with known magnetic moments in the excited states,
which satisfy
the  single-quasiparticle approximation and
give an additional possibility to check our main
approximation of no  interaction between two quasi-particles. The
results are given in Table 8 and they again confirm the model used.

\begin{table*}[tbp]
\caption{Magnetic moments ($\mu_N$) of odd-odd semimagic nuclei  in excited states. }
\begin {tabular}{l c c c c c c c c}
\hline
   Nucleus& $\lambda_1^n\lambda_2^p$ & $J^{\pi}$  &$T_{\rm 1/2}$& Level, keV& $\mu_{\rm phen}$ &
   $\mu_{\rm th}$ & $\mu_{\rm exp}$ \cite{stone,BNL} \\
\hline
${^{60}_{27}}$Co$_{33}$ &(2p$_{3/2}^n$ 1f$_{7/2}^p$)& $2^+$ &10.467 min&1121.36&+4.466&+4.687&+4.40 (9) \\

${^{90}_{41}}$Nb$_{49}$ & (1g$_{9/2}^n$ 1g$_{9/2}^p$) & $6^+$ & 63 $\mu$s&298.22&&+3.370&+3.720 (24)\\

${^{138}_{55}}$Cs$_{83}$ & (2f$_{7/2}^n$ 1g$_{7/2}^p$)& $6^-$ &2.91 min&721&+1.603&+1.455&+1.713 (9) \\

${^{138}_{57}}$La$_{81}$& (2d$_{3/2}^n$ 1g$_{7/2}^p$) & $3^+$ &116 ns&424&+2.385&+2.419&+2.886 (48) \\

${^{142}_{59}}$Pr$_{83}$& (2f$_{7/2}^n$ 2d$_{5/2}^p$) & $5^-$ &14.6 min&937.20&+2.329&+2.345&2.2 (1) \\
\hline

\end{tabular}
\end{table*}

\subsubsection{The case of strong configuration mixture: Tl isotopes and others.}

In this subsection, we consider several cases of odd-odd semimagic nuclei
where it is necessary go beyond our simple model because of the strong configuration
mixture of due to close single-particle energies in one of the odd subsystems. Results are
presented in Table 9. Seven odd-odd Tl
isotopes (A = 192-204) present the most instructive example. The experimental values of their magnetic moments
are small, see Table 9, and change the sign in the middle of the chain. It is a signal that
the low-lying neutron single-particle states $p_{3/2}$ and $f_{5/2}$ participate on
approximately equal footing in the odd-odd wave function. Indeed, as one can see from
Table 2, the magnetic moment values in odd Pb isotopes in these states have opposite signs and
approximately equal absolute values. Therefore, their partial contributions to the magnetic
moment of the odd-odd isotopes should compensate each other.
In our single-particle  self-consistent
scheme these levels are almost  degenerated,  so
the realization of the  configuration
[$3s_{1/2}^p 3p_{3/2}^n$ + $3s_{1/2}^p 2f_{5/2}^n$]
 can be realistic in order to explain the small values of magnetic moments in Tl
isotopes. Explicit theory of the configuration mixing is
 rather complicated, therefore, at this stage we construct the odd-odd wave function
 using a fitting procedure.
So, we introduce the corresponding mixing coefficient $\alpha$ and find it
to explain the experimental data:

\vspace{0.3cm}

\beq
\psi = \alpha\psi^{(1)}+\sqrt{1-\alpha^2}\psi^{(2)},
\eeq
\beq
\mu=\alpha^2\mu^{(1)}+(1-\alpha^2)\mu^{(2)},
\eeq
where $ \psi^{(1)}=$[$3s_{1/2}^p 3p_{3/2}^n$] and $ \psi^{(2)}=$[$3s_{1/2}^p 2f_{5/2}^n$],
$\mu^{(1)}$ and $\mu^{(2)}$ are magnetic moments of the configurations (1) and (2).
 It turned out that  the  $\alpha$ values are 0.60-0.77, with a smooth $A$-dependence.
Approximately, they  are described by the formula $\alpha = 0.5 - 3.18\left(\dfrac{A}{210.11}-1 \right) $.
As can be seen from Table 9,
the configuration mixture caused by almost degenerated two neutron single-particle levels
does explain small values of  magnetic moments of odd-odd Tl isotopes including the change
of the sign in the middle of the chain.

A similar situation with close position of single-particle levels there
is in ${^{58}_{27}}$Co$_{31}$ , ${^{106}_{49}}$In$_{57}$,
${^{110}_{49}}$In$_{61}$
and ${^{124}_{51}}$Sb$_{73}$ nuclei, where
 we have mixed the following configurations:

${^{58}_{27}}$Co$_{31}$ : $\alpha[1f_{7/2}^p 2p_{3/2}^n]$+$\sqrt{1-\alpha^2}[1f_{7/2}^p 1f_{5/2}^n]$,

${^{106}_{49}}$In$_{57}$ : $\alpha[1g_{9/2}^p 1g_{7/2}^n]$+$\sqrt{1-\alpha^2}[1g_{9/2}^p 2d_{5/2}^n]$,

${^{110}_{49}}$In$_{61}$ : $\alpha[1g_{9/2}^p 1g_{7/2}^n]$+$\sqrt{1-\alpha^2}[1g_{9/2}^p 2d_{5/2}^n]$,

${^{124}_{51}}$Sb$_{73}$ : $\alpha[1g_{7/2}^p 1h_{11/2}^n]$+$\sqrt{1-\alpha^2}[1d_{5/2}^p 1h_{11/2}^n]$.

The coefficients $\alpha$ obtained by the simple fitting procedure are given in Table 9. They give
us the information about the quasiparticle structure of the odd-odd nuclei under consideration where the
simple model we used previously does not work. It can be used for finding other characteristics of these
nuclei.

In our single-particle scheme, we have found nuclei with close energies
of three (in $^{88}$Rb, $^{90}$Y) and four  (in $^{114}$Sb, $^{116}$Sb) single-particle levels.
To reliably describe such nuclei, the explicit theory of configuration mixing should be developed.


\begin{table}[b]
\caption{Ground state magnetic moments ($\mu_N$)  of odd-odd Tl isotopes  and ${^{58}_{27}}$Co$_{31}$ , ${^{106}_{49}}$In$_{57}$
${^{110}_{49}}$In$_{61}$
and ${^{124}_{51}}$Sb$_{73}$ obtained by mixing two configurations ($\alpha$ is the fitted mixing coefficient).}
\begin {tabular}{l c c c c c c c c}
\hline
   Nucleus  &$T_{\rm 1/2}$ & $J^{\pi}$ &$\alpha$&  $\mu_{\rm th}$ & $\mu_{\rm exp} $\cite{stone,BNL} \\
\hline
${^{58}_{27}}$Co$_{31}$& 70.86 d& $2^+$ &0.806&+4.044&+4.044 (8)\\

${^{106}_{49}}$In$_{57}$  & 6.2 min& $7^+$ &0.450&+4.916&4.916(6)\\

${^{110}_{49}}$In$_{61}$  & 4.9 h& $7^+$ &0.373&+4.713&+4.713 (8)\\


${^{124}_{51}}$Sb$_{73}$  &60.20 d& $3^-$ &0.692&-1.20&-1.20 (2)\\

${^{192}_{81}}$Tl$_{111}$  &9.6 min & $2^-$ &0.774& +0.199 &+0.200 (3) \\

${^{194}_{81}}$Tl$_{113}$  &33.0 min& $2^-$ &0.744& +0.142 &0.140 (3)\\

${^{196}_{81}}$Tl$_{115}$ &1.84 h& $2^-$ &0.714& +0.057 &0.072 (3)\\

${^{198}_{81}}$Tl$_{117}$  &5.3 h& $2^-$ &0.683& +0.020 &0.00 (1)\\

${^{200}_{81}}$Tl$_{119}$  &26.1 h& $2^-$ &0.653& -0.038 &0.04 (1)\\

${^{202}_{81}}$Tl$_{121}$  &12.31 d& $2^-$ &0.623& -0.069 &0.06 (1)\\

${^{204}_{81}}$Tl$_{123}$ &3.783 y& $2^-$ &0.593& -0.089  &0.09 (1)\\

\hline
\end{tabular}
\end{table}

\section{Comparison with the MPSM calculations \cite{BABr}.  }

In Table 10, we compare our results $\mu_{\rm th}$ for odd and odd-odd nuclei with those
obtained within the
MPSM approach \cite{BABr} for  $fp$ shell nuclei.
We did not  include in this table the $^{58}$Co nucleus for which,
see Table 9, agreement is obtained by fitting the coefficient $\alpha$.
The agreement
between the experiment,  our results and the usual shell model calculations $\mu_{\rm th1}$ are
rather reasonable.  The improved shell model results   $\mu_{\rm th2}$, which were obtained
due to using some new fitted parameters,
 give a little better agreement with the experiment. It should be noted that in our calculations
 there were no new or fitted parameters and our method is applicable to any nuclei, not only the $fp$ shell ones.

\begin{table}[b]
\caption{Comparison of our model with the shell model  calculations $\mu_{\rm th 1}$ and $\mu_{\rm th 2}$ for the $fp$ shell nuclei \cite{BABr}}
\begin {tabular}{l c c c c }
\hline
   Nucleus  &$\mu_{\rm exp} $ \cite{stone,BNL} &   $\mu_{\rm th}(TFFS)$ & $\mu_{\rm th 1}$&$\mu_{\rm th 2}$ \\
\hline
${^{55}_{27}}$Co$_{28}$&+4.822 (3)&+4.793&+4.630&+4.746\\

${^{56}_{27}}$Co$_{29}$& 3.851 (12)&+3.949&+3.652&+3.774\\

${^{57}_{27}}$Co$_{30}$&+4.720 (10)&+4.899&+4.616&+4.704\\

${^{59}_{27}}$Co$_{32}$&+4.627 (9)&+4.609&+4.637&+4.707\\

${^{60}_{27}}$Co$_{33}$&+3.799 (8)&+3.993&+3.962&+3.996\\

${^{57}_{28}}$Ni$_{29}$&-0.7975 (14)&-0.812&-0.789&-0.802\\

${^{61}_{28}}$Ni$_{33}$&-0.75002 (4)&-0.829&-0.688&-0.707\\

\hline
\end{tabular}
\end{table}

\section{Conclusion}
A simple model of no  interaction between the odd proton and the odd
neutron is used to describe magnetic moments of odd-odd nuclei in terms of the
magnetic moments of the neighboring odd nuclei. To check this approximation,
a phenomenological analysis was carried out where the experimental magnetic
moments of odd nuclei were used. It showed that in the major number of nuclei
the model works reasonably well.

The magnetic moments of about one-hundred  odd and more than one-hundred
odd-odd nuclei have been calculated within the universal approach
which uses the self-consistent single particle basis calculated with
the EDF by Fayans {\it et al.} and the TFFS effective spin forces. The
parameters of the EDF and spin forces are well-known and have been
successfully used in the calculations of many other nuclear
characteristics. The experimental data for about sixty  new (as
compared with Ref. \cite{mu1}) odd nuclei were reproduced with almost
the  same  accuracy of 0.1-0.2$\mu_N$ as in the previous work
\cite{mu1}. The experimental values of magnetic moments  for about 20
odd-odd semimagic nuclei in ground and excited states
were reproduced with approximately the same accuracy.
The values of magnetic moments for 60 odd-odd magic nuclei in
excited states have been predicted.
Several odd-odd nuclei are considered including seven Tl isotopes,
where it is necessary to go beyond our simple model and take into account
the configuration mixture. Agreement with the data is automatically obtained by
fitting the mixing coefficient. It can be used for calculations of other characteristics
of these nuclei.

In all cases where the experimental signs of magnetic moments of odd and odd-odd nuclei are known,
our calculations gave the same signs, therefore, our predictions of the signs  may be considered as reliable.
Most of considered odd and odd-odd nuclei are unstable with a rather short  time of life.
In our calculations, we have used a small number of known  parameters, which are universal and can be used
for all nuclei. For this reason, our predictions for unstable nuclei may be considered as  rather reliable.

\vspace{0.5cm}
\section{Acknowledgment}
The work was partly supported by the DFG and RFBR Grants
Nos.436RUS113/994/0-1 and 09-02-91352NNIO-a, by the Grants
NSh-215.2012.2, NSh-7235.2010.2  and 2.1.1/4540 of the Russian
Ministry for Science and Education, and by the RFBR grants
11-02-00467-a, 12-02-00955-a, 13-02-00085-a, and 13-02-12106.
We are grateful to Institut f\"ur Kernphysik, FZ
J\"ulich for hospitality.

{}

\end{document}